\documentclass[fleqn,usenatbib]{mnras}

\usepackage{newtxtext,newtxmath}
\usepackage[T1]{fontenc}

\DeclareRobustCommand{\VAN}[3]{#2}
\let\VANthebibliography\thebibliography
\def\thebibliography{\DeclareRobustCommand{\VAN}[3]{##3}\VANthebibliography}

\usepackage{bm}
\usepackage{graphicx}	% Including figure files
\usepackage{amsmath}	% Advanced maths commands
\title[Light-cone effect on \ion{C}{ii} signal statistics]{\ion{C}{ii} and \ion{H}{i} 21-cm line intensity mapping from the EoR: Impact of the light-cone effect on auto and cross-power spectra}

\author[C. S. Murmu et al.]{Chandra Shekhar Murmu,$^{1}$\thanks{E-mail: chandra0murmu@gmail.com (CSM)}
Suman Majumdar$^{1,\, 2}$
and Kanan K. Datta$^{3}$
\\
$^{1}$Department of Astronomy, Astrophysics and Space Engineering, Indian Institute of Technology Indore, Khandwa Rd., Simrol, MP 453552, India\\
$^{2}$Department of Physics, Blackett Laboratory, Imperial College, London SW7 2AZ, U. K.\\
$^{3}$Department of Physics, Presidency University, 86/1 College St., Kolkata 700073, India\\}

\date{Accepted 2021 August 10. Received 2021 August 7; in original form 2021 July 16}

\pubyear{2021}

\begin{document}
\label{firstpage}
\pagerange{\pageref{firstpage}--\pageref{lastpage}}
\maketitle

% Abstract of the paper
\begin{abstract}
\ion{C}{ii} line intensity mapping (LIM) is a potential technique to probe the early galaxies from the Epoch of Reionization (EoR). Several experiments e.g. CONCERTO, TIME, CCAT-p are underway to map the \ion{C}{ii} LIM signal fluctuations from the EoR, enabling us to estimate the \ion{C}{ii} power-spectrum and \ion{C}{ii}$\times$21-cm cross-power spectrum. Observed LIM signal will have its time evolution embedded in it along the Line of Sight (LoS) due to the finite travel time of the signal from its origin to the observer. We have investigated this so-called light-cone effect on the observed statistics of our semi-numerically simulated \ion{C}{ii} signal from the EoR. Using a suit of simulated \ion{C}{ii} and neutral hydrogen 21-cm maps and corresponding light-cone boxes, we have shown that the light-cone effect can impact the \ion{C}{ii} power spectrum by more than 15\% at large scales ($k\sim 0.1\, \text{Mpc}^{-1}$, at $z=6.8$). We have also observed that the impact of light-cone effect on the \ion{C}{ii} power spectrum drops with decreasing redshift within the redshift range considered here ($7.2 \lesssim z \lesssim 6$). The \ion{C}{ii}$\times$21-cm cross-power spectrum is also affected by light-cone, and in our models where reionization ends before $z=6$, we find that the maximum impact on cross-power can reach up to 20\%. At $z=6.4$, we find comparatively pronounced variation in the light-cone effect with reionization history on the cross power. Faster reionization histories have a more drastic light-cone effect on cross-power. We conclude that we need to incorporate the light-cone in order to properly model the signal,   constrain the EoR-related astrophysical parameters and reionization history using the \ion{C}{ii}$\times$21-cm cross-power spectrum.
\end{abstract}

% Select between one and six entries from the list of approved keywords.
% Don't make up new ones.

%%%%%%%%%%%%%%%%%%%%%%%%%%%%%%%%%%%%%%%%%%%%%%%%%%

%%%%%%%%%%%%%%%%% BODY OF PAPER %%%%%%%%%%%%%%%%%%

\section{Introduction}
\label{Intro}
Understanding the mysterious Epoch of Reionization (EoR) has been at the forefront of modern cosmology research. The first light sources appeared in the universe during the EoR, and subsequently reionized the neutral inter-galactic medium (IGM) with ionizing radiation. The current constraints on the EoR come from, among the others, the Gunn-Peterson trough in the very distant quasar spectra \citep[]{Becker_2001, Becker_2015}, the constraint on the Thomson optical depth ($\tau = 0.054 \pm 0.007$) for instantaneous reionization model \citep[]{Plank_2020} and detection of the global-averaged 21-cm signal from Cosmic Dawn by EDGES \citep[][]{Bowman_2018}. Thus the pressing question about the EoR - how did the reionization process take place, still remains largely unknown. Critical questions like: how did the reionization history proceed?; what are the properties of the sources responsible for reionization?; can we shed light on the history of structure formation by probing the EoR?; also remain unanswered. We need direct observations that can probe the evolving state of the early IGM and the early reionizing sources during the EoR.

State-of-the-art radio interferometers like LOFAR \citep[]{Mertens_2020}, MWA \citep[]{barry19, Patwa2021}, PAPER \citep[]{kolopanis19}, HERA \citep[]{deboer17}, GMRT \citep[]{Choudhuri_2020, Pal2021} etc. are trying to detect the spatial fluctuations in the 21-cm signal originating from the neutral hydrogen (\ion{H}{i}) of the early IGM during the EoR. These facilities are also acting as the technology and science precursors or path-finders for the upcoming humongous Square Kilometre Array (SKA)\footnote{\url{https://astronomers.skatelescope.org}}, the largest radio telescope to be ever built by humankind. The SKA's contribution via production of first high resolution tomographic images of the 21-cm signal from the EoR will be a significant milestone in this paradigm \citep[]{koopmans15, mellema15}. Though many of the existing radio interferometric experiments have advanced significantly, removing of $5-6$ orders of magnitude larger foreground contamination from  21-cm observations remains the major challenge \citep[e.g.][]{dimatteo02, ali08, jelic08, ghosh12, Procopio_2017} in detecting this signal from the EoR. The ongoing experiments thus have been able to produce only weak upper limits to the signal power spectrum at a few redshifts \citep{paciga13, Mertens_2020, barry19, li19, kolopanis19, trott20,  Pal2021}.
Once the \ion{H}{i} 21-cm power spectrum across different redshifts are detected one might be able to constrain the IGM physics, reionization history and, most likely, one can also put some indirect constraints on the reionization source properties. However, the 21-cm observations alone are not enough to get a complete picture of the EoR.

To have a comprehensive understanding of the EoR one would need to probe the sources of the reionizing photons, i.e. the early galaxies and quasars etc., along with the state of the IGM. However, the spectroscopic detection of individual galaxies from the EoR is challenging. An ideal choice for doing cosmological large-volume surveys of early sources of the EoR is to use the line intensity mapping technique \citep[]{Visbal_2010,Gong_2011_2,Gong_2011,Silva_2013,Silva_2015}. In the line intensity mapping (LIM), one does not need to resolve the individual light sources but instead need to detect the integrated flux from several sources  which lies within a pixel or voxel of the 3D volume that is being surveyed by the telescope. It targets a specific atomic or molecular line transition from galaxies and uses it as a tracer to detect them \citep[]{Fonseca_2016} via this method. The ideal line emission choices are the ones that are otherwise bright and can provide valuable information regarding various properties  of those galaxies.

The \ion{C}{ii} $158\, \umu$m line is expected to be one of the brightest line emissions from the early galaxies and this line emission also correlates well with the star-formation rate within a galaxy \citep[][]{De_Looze_2011}. The recent spectroscopic surveys have detected galaxies at high-redshifts using this line emission. The ALPINE-ALMA \ion{C}{ii} survey has detected star-forming galaxies at $4<z<6$ and was also able to classify them \citep[A2C2S,][]{Le_F_vre_2020}. The ALMA Spectroscopic Survey in the Hubble Ultra Deep Field (ASPECS) also searched for potential \ion{C}{ii} emitters at $6<z<8$. They were able to place upper limits on the \ion{C}{ii} luminosity and the cumulative number density of \ion{C}{ii} emitters \citep[][]{Uzgil_2021} in that redshift range. These high-redshift detections have further strengthened confidence in the upcoming missions like e.g. CONCERTO \citep[][]{lagache_2017,Concerto}, the CCAT-p \citep[][]{Cothard_2020}, and TIME \citep[][]{Crites_2014, Sun_2021}, which will be conducting \ion{C}{ii} LIM and probe the spatial fluctuations in \ion{C}{ii} emission from early galaxies at different redshifts.

By measuring the fluctuations in the \ion{C}{ii} line intensity maps from the early galaxies one would expect to obtain valuable physical insights about the properties of these sources and their role in reionization. By far, the most common statistic that has been considered for measuring this signal fluctuations, is the power spectrum. The power spectrum measures the amplitude of fluctuations in any signal at different length scales. Combining these information together one can constrain the role of early galaxies in reionization. The \ion{C}{ii} power spectrum can in principle tell us about the \ion{C}{ii} luminosity function \citep[]{Yue_2019} and its redshift evolution. Additionally, it can also help us to quantify the clustering properties of the ionizing galaxies. Several recent studies have forecasted that the \ion{C}{ii} power spectrum should be detectable with experiments like CONCERTO up to $z=8$; stage-II experiments similar to CCAT-p can push towards higher redshifts with greater signal-to-noise ratio (SNR) \citep[]{Dumitru_2019}. 

On the other hand, the cross-power spectrum of the \ion{C}{ii} and  21-cm signals from the same redshifts is another statistic of particular interest \citep[]{Chang_2015}. Reliable detection of the auto-power spectrum of either 21-cm or \ion{C}{ii} signals depends on successful removal of the foregrounds emissions. However, in  the cross-power spectrum of the \ion{C}{ii} and  21-cm signals, foregrounds in the individual maps are expected to be  uncorrelated to each other. On the other hand one would expect the \ion{C}{ii} and 21-cm signals to be anti-correlated at large length scales and would expect their level of anti-correlation to vary with length scales and the stage of reionization. This essentially makes the  detection of both the signal via a cross-correlation statistics e.g. the cross-power spectrum more feasible compared to their auto power spectrum. However, one has to keep in mind that the foregrounds will still contribute in the errors of the signal statistics even when it is detected via cross-correlation.
Once detected, one in principle can use the \ion{C}{ii}$\times$21-cm cross-power spectrum to constrain the reionization history and the EoR parameters \citep[]{Dumitru_2019} more precisely. 

Any electromagnetic radiation takes a finite amount time to travel from its point of origin to the observer. When any cosmological signal originating at different distances along the line-of-sight (LoS) of a present day observer, arrives at the present day observer at the same time, the signal and its statistics will change along the LoS. This is due to the fact that signals coming from different cosmic distances were originated at different cosmic times. This effect is popularly known as the `light-cone effect'. As one can anticipate such a light-cone effect is expected to have an impact on the estimated signal statistics, such as on the power spectrum and cross-power spectrum etc. In the context of the EoR, it has been demonstrated that this affects the \ion{H}{i} 21-cm signal two-point correlation function and power spectrum \citep[]{Datta_2012,Datta_2014,Zawada_2014, La_Plante_2014,Ghara_2015, mondal18}. One would expect the light-cone effect to have a similar impact on the \ion{C}{ii} power spectrum and the \ion{C}{ii}$\times$21 cm cross-power spectrum. However, as per our knowledge, the impact of the light-cone effect on the \ion{C}{ii} power spectrum and the \ion{C}{ii}$\times$21 cm cross-power spectrum has not been quantified so far.

In this article we aim to quantify the light-cone effect on the \ion{C}{ii} power-spectrum and \ion{C}{ii}$\times$21 cm cross-power spectrum from the EoR. It is important to quantify this effect on the expected signal, as it can potentially affect the predicted detectability of the signal. Additionally, if this effect is not properly taken into account that may lead to wrong interpretation of observed signal . To quantify this effect, we have first simulated both the \ion{C}{ii} and the 21-cm signals from the EoR. We have used a combination of N-body dark matter only simulation, a Friend-of-Friend halo finder and a set of semi-numerical prescription for simulating the EoR 21-cm and \ion{C}{ii} intensity maps for a series of fixed redishifts (coeval maps). We then have constructed light-cone cuboids of the simulated signals from these redshift snapshots. While simulating and constructing the light-cone cuboids for the 21-cm signal, we also consider different reionization histories to have varied light-cone impact on the simulated signal. We next estimate the spherically averaged auto power spectrum of the \ion{C}{ii} and 21-cm signals from both the coeval cubes and light-cone cubes and compare them to asses the impact of the light-cone effect on these two different signals. Next, we have studied  the light-cone effect on the \ion{C}{ii}$\times$21-cm cross-power spectrum.

This paper is structured as follows. In section \ref{Signal}, we briefly describe the signals that we are interested in studying, and then we elaborate on the simulation methods of these signals in section \ref{Sim}. In section \ref{Result}, we define the power-spectrum estimator and discuss our findings. Finally, we summarise the paper in section \ref{Summary}.

Throughout this work, we have adopted cosmological parameters $\Omega_\text{m}=0.3183,\Omega_\Lambda=0.6817,h=0.6704,\Omega_\text{b}h^2=0.022032,\sigma_8=0.8347,n_\text{s}=0.9619$, consistent with Planck+WP best-fit values \citep[]{Planck_XVI}.

\section[The EoR C II \texorpdfstring{158$\umu$m}{} and H I 21 cm line intensity mapping signals]{The E\lowercase{o}R C\,{\sevensize II} \texorpdfstring{158$\bmath{\umu}$\lowercase{m}}{} and H\,{\sevensize I} 21-\lowercase{cm} line intensity mapping signals} \label{Signal}

The \ion{C}{ii} 158$\umu$m line is originated from singly ionized carbon atoms,  which results from the fine-structure transition between $^2P_{3/2} \rightarrow ^2P_{1/2}$. Although it is believed that \ion{C}{ii} emission originates from the photo-dissociation regions (PDR) of galaxies, a significant amount is also generated from the cold neutral medium, warm ionized medium \citep[]{De_Looze_2011}. However, it is expected that the \ion{C}{ii} emission from the ISM of galaxies to be more intense as compared to the emission from the IGM \citep[]{Gong_2011}, therefore it can be treated as a good tracer of the galaxies and star formation.

The intensity of the \ion{C}{ii} emission at 158$\umu$m can be modelled by assuming \ion{C}{ii} luminosity ($L_\ion{C}{ii}(M,z)$) for a given halo with mass $M$ at redshift $z$, residing within a co-moving volume $\Delta V$. Then the \ion{C}{ii} LIM signal can be expressed as,
\begin{equation}
    I_\text{\ion{C}{ii}} = \frac{1}{\Delta V}\sum_{i} \frac{L_{\text{\ion{C}{ii}}}(M_i,z)}{4\pi D_\text{L}^2 }\frac{d\chi}{d\nu_{\text{obs}}}D_\text{A}^2 \,,
    \label{eq:1}
\end{equation}
with $\chi$ and $\nu_{\mathsf{obs}}$ being the co-moving distance and observed frequency, respectively. The $D_\text{L}$ and $D_\text{A}$ represent the proper luminosity-distance and comoving angular-diameter distance, respectively. Therefore, we can rewrite $I_\text{\ion{C}{ii}}$ as \citep[][]{Visbal_2010, Silva_2015},
\begin{equation}
    I_\text{\ion{C}{ii}} = \frac{1}{\Delta V} \sum_{i} \frac{L_\text{\ion{C}{ii}}(M_{i},z)}{4\pi H(z)}\lambda_\text{\ion{C}{ii}} \,,
    \label{eq:2}
\end{equation}
and the mean of the signal is expressed as,
\begin{equation}
    \Bar{I}_\text{\ion{C}{ii}} = \int_{M_{\mathsf{min}}}^{M_{\mathsf{max}}} dM \frac{dn}{dM} \frac{L_\text{\ion{C}{ii}}(M,z)}{4\pi H(z)}\lambda_\text{\ion{C}{ii}}\,,
    \label{eq:3}
\end{equation}
with $\lambda_\text{\ion{C}{ii}}$ being the rest-frame wavelength of \ion{C}{ii} emission and $dn/dM$ is the halo-mass function.

The \ion{H}{i} 21-cm signal originates from the transition between the hyperfine states of the ground state neutral hydrogen. During the EoR, most of the neutral hydrogen gas is present in the IGM, so the 21-cm signal from the EoR essentially traces the diffuse gas density of the early IGM. This signal from the EoR can be detected as an excess brightness temperature relative to the CMBR, which is expressed as \citep[][]{Bharadwaj_2005},
\begin{equation}
    T_b = 4\text{mK}\, \Bar{x}_{\text{\ion{H}{i}}} (1+\delta_{\text{\ion{H}{i}}})\bigg(\frac{\Omega_\text{b} h^2}{0.02}\bigg)\bigg(\frac{0.7}{h}\bigg)\sqrt{\frac{1+z}{\Omega_\text{m}}}\,,
    \label{eq:4}
\end{equation}
where $\Bar{x}_\text{\ion{H}{i}}$ is the mass-averaged neutral hydrogen fraction and $\delta_{\text{\ion{H}{i}}}$ is the neutral hydrogen over-density and the rest of the quantities are the cosmological parameters. The effects of \ion{H}{i} peculiar velocities are neglected in this expression and we also assume that the $T_\text{s} \gg T_\gamma$ during the EoR, where $T_\text{s}$ is the spin temperature of the neutral hydrogen gas and $T_\gamma$ is the temperature of the Cosmic Microwave Background Radiation (CMBR). In the next section, we discuss about the methods that we have adopted to simulate these two signals from the  EoR.

\section{Simulations} \label{Sim}
We employ a particle-mesh (PM) N-body\footnote{\url{https://github.com/rajeshmondal18/N-body}} \citep[]{bharadwaj04} code that generates dark matter density fields at required redshifts. Our simulation has a volume of $215^3$ cMpc$^3$ with $3072^3$ grids and $1536^3$ particles. It results in a grid resolution of $0.07$ cMpc and particle-mass resolution of $\approx 10^8 M_\odot$. We then use a Friends-of-Friends (FoF)\footnote{\url{https://github.com/rajeshmondal18/FoF-Halo-finder}} \citep[]{Mondal_2015} algorithm to identify collapsed halos. The linking length used for this is $0.2$ times the mean inter-particle separation in the simulation. The resulting halo-mass resolution is $\approx 10^9 M_\odot$, and we generate snapshots at $13$ redshifts between $z = 6$ and $7.2$, separated by $\Delta z=0.1$. We then have generated the \ion{C}{ii} and the \ion{H}{i} 21-cm maps out of these simulation snapshots. The grid resolution for these maps are reduced from $3072^3$ to $384^3$ and grid separation is raised to $0.56$ cMpc. All the volumes and length-scales reported in this work are in comoving units, and hereafter we drop the cMpc notation in favour of Mpc.

\subsection[CII maps]{C\,{\sevensize II} maps}
\begin{figure}
    \centering
    \includegraphics[width=\columnwidth]{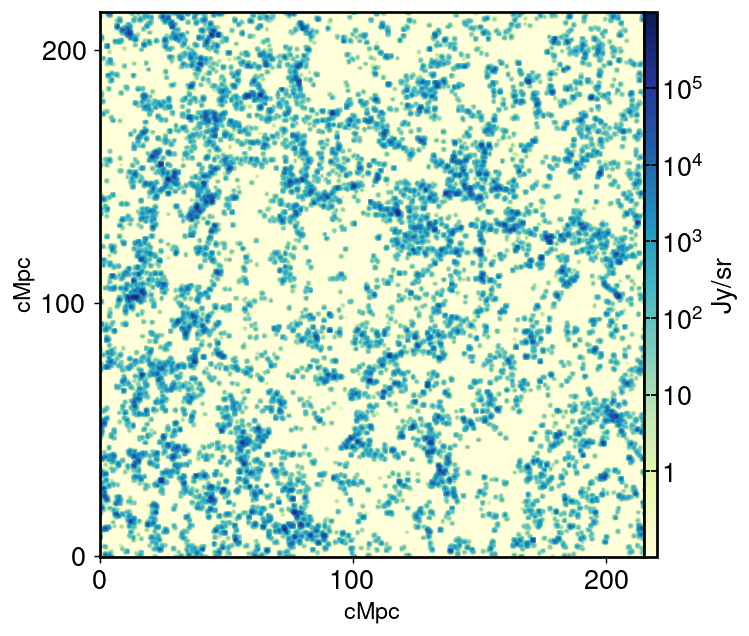}
    \caption{This figure shows a slice of \ion{C}{ii} map at $z=7.2$.}
    \label{fig:CII_map}
\end{figure}

The \ion{C}{ii} maps are generated by painting \ion{C}{ii} luminosity to the halos identified in the dark matter distribution generated by the N-body simulation. To perform this we utilize a set of parametric relationships to connect the halo mass to the \ion{C}{ii} luminosity. First, the halo mass is linked to the star-formation rate (SFR) of that halo using
\begin{align}
    \frac{\text{SFR}}{M_{\sun} yr^{-1}}=2.25\times10^{-26}\big(1+(z-7)7.5\times10^{-2}\big) \notag \\
    \times M^a\bigg(1+\frac{M}{c_1}\bigg)^b\bigg(1+\frac{M}{c_2}\bigg)^d\bigg(1+\frac{M}{c_3}\bigg)^e
    \label{eq:5}
\end{align}
with $a=2.59$, $b=-0.62$, $d=0.4$, $e=-2.25$, $c_1=8\times10^8 M_{\sun}$, $c_2=7\times10^9 M_{\sun}$, $c_3=10^{11} M_{\sun}$, adopted from \citet[eq. 26]{Silva_2013}. We then relate the SFR to the \ion{C}{ii} luminosity for that halo using
\begin{equation}
    \log_{10}\bigg(\frac{L_\text{\ion{C}{ii}}}{L_{\sun}}\bigg) = a_{L_\text{\ion{C}{ii}}}\times \log_{10}\bigg(\frac{\text{SFR}}{M_{\sun} yr^{-1}}\bigg) + b_{L_\text{\ion{C}{ii}}}
    \label{eq:6}
\end{equation}
\citep[see][eq. 7]{Silva_2015}. Based on this prescription we generate a \ion{C}{ii} luminosity distribution from the simulated dark matter halo distribution. Next, using the Cloud in Cell method, we map this distribution to a coarse gridded map of \ion{C}{ii} luminosity. Then for each grid point, we determine the \ion{C}{ii} intensity using equation \eqref{eq:2}. This finally gives us the \ion{C}{ii} intensity map for each redshift as shown in Fig.~\ref{fig:CII_map}.

The choice of parameters for equation~\eqref{eq:6} adopted throughout this work, corresponds to the model $\textit{m1}$ ($a_{L_\text{\ion{C}{ii}}}=0.8475,b_{L_\text{\ion{C}{ii}}}=7.2203$) of \citet[]{Silva_2015}. This choice leads to the highest value of $\Bar{I}_\ion{C}{ii}$ among the four models. Our model of \ion{C}{ii} signal does not account for the population abundance of high-redshift \ion{C}{ii} emitters. This adoption corresponds to the most optimistic scenario, where every halo is an efficient emitter of \ion{C}{ii} line. Due to this, the $\Bar{I}_\ion{C}{ii}$ in this work is higher by a factor of $\sim 2$ compared to that of \citet{Silva_2015} for the model $\textit{m1}$. Nonetheless,  the analysis of light-cone remains unaffected without any loss of generality.

\subsection[HI maps]{H\,{\sevensize I} maps}
\label{sec:HI_maps}
The \ion{H}{i} maps are generated using the semi-numerical code ReionYuga\footnote{\url{https://github.com/rajeshmondal18/ReionYuga}} \citep[]{Choudhury_2009,Majumdar_2014,Mondal_2015}, which is based on excursion-set formalism \citep[]{Furlanetto_2004}. The \ion{H}{i} distribution follows the underlying dark matter distribution in our simulation. Free parameters used to tune the reionization history are the minimum mass of dark matter halos contributing to reionization ($M_{\text{h-min}}$), ionizing photon-emitting efficiency ($N_{\text{ion}}$), and ionizing photon mean-free-path ($R_{\text{mfp}}$). $M_{\text{h-min}}$ sets the minimum limit of halo mass below which no halos are considered for contribution to reionization. The number of ionizing photons emitted by a source is proportional to the halo mass of the halo hosting them in our simulation. The proportionality constant here is the $N_{\text{ion}}$. The  $R_{\text{mfp}}$ is used as the maximum radius to smooth the hydrogen and photon density fields and determine the ionization condition at grid points. We start with a smoothing radius equal to grid separation and end at $R_{\text{mfp}}$ to check whether at any radius the smoothed photon density exceeds the smoothed neutral hydrogen density at a given grid point. If the condition is met, the grid point is reckoned to be ionized.

\begin{figure}
    \centering
    \includegraphics[width=\columnwidth]{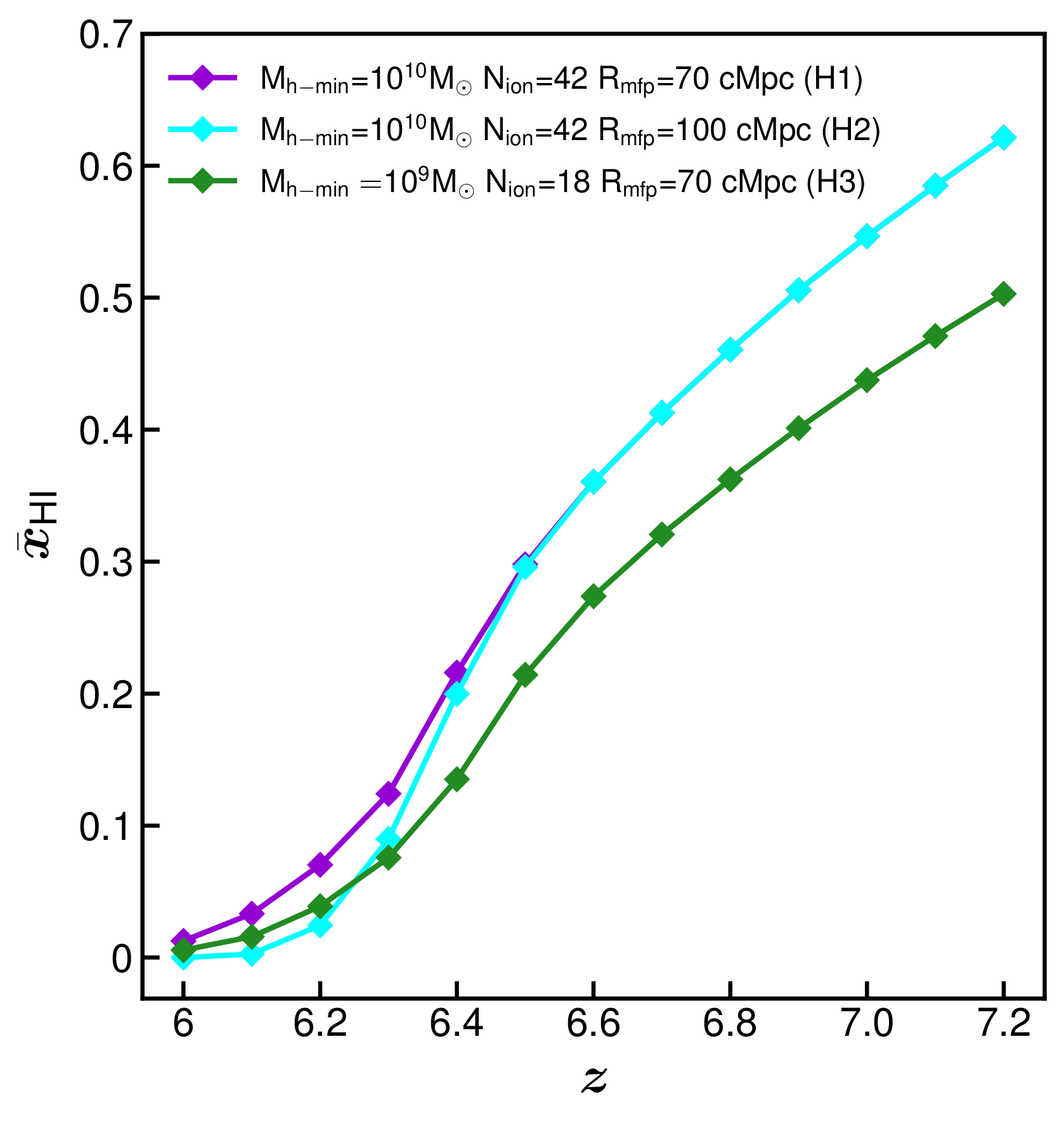}
    \caption{Neutral fraction ($x_{\mathsf{\ion{H}{i}}}$) vs. Redshift ($z$) for three different reionization histories.}
    \label{fig:xHI_vs_z}
\end{figure}

By tuning these free parameters, we have generated three different reionization histories (H1, H2, H3, Fig.~\ref{fig:xHI_vs_z}), each corresponding to a set of the EoR parameters. These reionization histories differ in the evolution of neutral fraction with redshift, but reionization finishes at $z \sim 6$ in all scenarios.

Depending on the reionization history, both the mean and the fluctuations in the \ion{H}{i} 21-cm signal will evolve differently. It means that for different reionization histories the light-cone effect will impact the \ion{H}{i} 21-cm power spectrum and the \ion{C}{ii}$\times$21-cm cross-power spectrum differently. One would expect that in a faster reionization history, the fast-evolving fluctuations will boost the impact of the light-cone effect. For example, in reionization history H2, having a higher $R_\text{mfp}$ means that ionized regions will develop faster than H1. In H3, having a lower minimum-halo-mass for ionizing photon emission means that the reionization gets boosted from these low-mass halos and leads to faster ionization bubble development. These factors play a vital role in determining the evolution of the \ion{H}{i} 21-cm signal fluctuations, which further contributes to the light-cone effect. In this work, we wish to demonstrate that for varying reionization histories how the impact of light-cone effect on the \ion{C}{ii}$\times$21cm cross-power spectrum also varies.

\subsection{Light-cone maps}
We generate light-cone maps using a formalism similar to \citet[]{Datta_2012}. The algorithm used here consists of the following basic steps:
\begin{enumerate}
    \item Given a set of coeval simulations and their redshifts, we choose $z_\text{low}$ and $z_\text{high}$ corresponding to the lowest and highest redshifts of the coeval boxes.
    \item If $l$ is the grid-spacing of the coeval boxes and $L$ is the co-moving distance between $z_\text{low}$ and $z_\text{high}$, we calculate the integer division $N=L/l$. A redshift list $\mathbb{Z}$ with $N$ entries $z_1,z_2,z_3,...,z_N (z_1 < z_2 < z_3 ... < z_N)$ is prepared, such that each redshift in the entry is separated by a co-moving distance $l$.
    \item If the number of slices in the coeval outputs are $M$ and we want to construct the $p$th slice of light-cone, we compute the remainder $q$ of the integer division $p/M$. We then take the $q$th slice of all the coeval outputs and using steffen's interpolation \citep[]{Steffen_1990}, construct the $p$th slice of light-cone at $z_p$ ($z_p \in \mathbb{Z}$).
\end{enumerate}

\begin{figure*}
    \centering
    \includegraphics[width=0.7\textwidth]{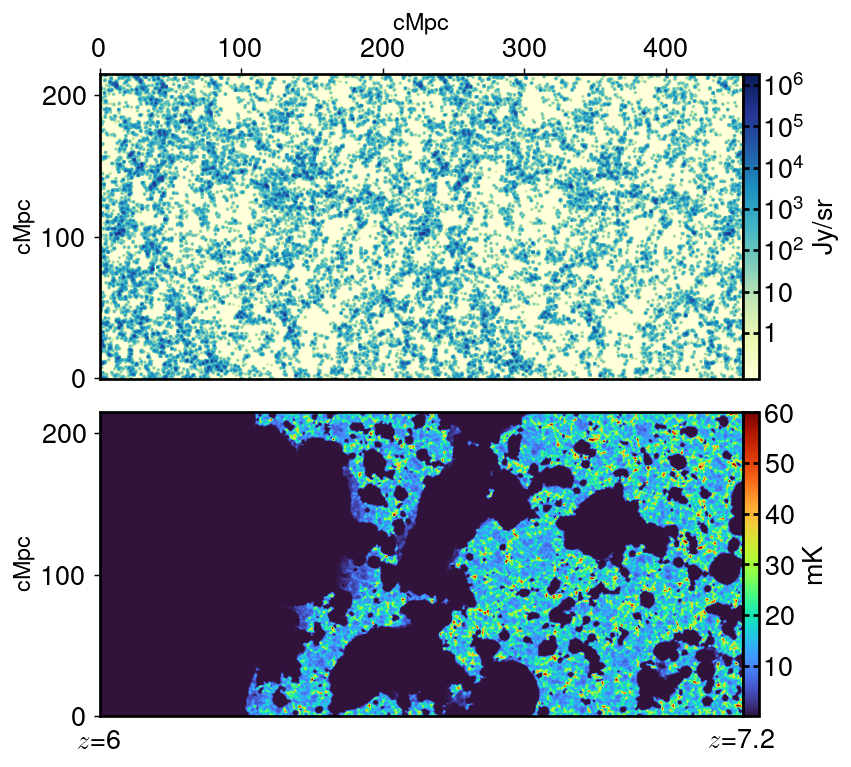}
    \caption{Light-cone slices for \ion{C}{ii} (top panel) and \ion{H}{i} 21-cm signal (bottom panel). The reionization history shown here is for model H1.}
    \label{fig:lc}
\end{figure*}

We apply this algorithm to generate both \ion{C}{ii} and \ion{H}{i} light-cone volumes. Slices from light-cones are shown in Fig.~\ref{fig:lc}. Our redshift range for coeval maps is $z=6$ to $7.2$, so the light-cone volumes also have the same redshift extent. The light-cone volumes have a full co-moving extent longer than the size of our coeval cubes. It leads to periodicity effects in the light-cone volumes, and in some cases, structures get repeated (e.g., see the \ion{C}{ii} light-cone in Fig.~\ref{fig:lc}). For this reason we cut out light-cone chunks of identical volume and size as the coeval cubes and compare their statistics. In some cases, the slices of light-cone volumes are rotated to get rid of the periodicity effects \citep[see][]{Zawada_2014}. In this work, we do not follow such a prescription. This method usually works with light-cone volumes of galaxies or discrete emitters, but not with the diffused gas e.g. \ion{H}{i} 21-cm light-cone maps. Since one of our principal goals is to study the cross-power spectrum, we need to keep the method for generating the light-cone volumes same for both signals. In terms of the statistics, since we cut out light-cone chunks of the exact size as the coeval cubes, we do not face the periodicity problem there and thus do not require to implement the corrective rotation trick. The percentage change in power-spectrum and cross-power spectrum of the signals  in light-cone is estimated with respect to the same from coeval volume, i.e. $\Delta (\%) = (lc - cc)\times100/cc$, where $\Delta$ is the percentage change, $lc$ corresponds to light-cone chunk power or cross-power spectrum and $cc$ corresponds to coeval cube. We show our light-cone analysis for the \ion{C}{ii} and \ion{H}{i} 21-cm power spectrum and the \ion{C}{ii}$\times$21-cm cross-power spectrum in the next section.

\section{Results} \label{Result}
Here we discuss the results of the impact of light-cone effect on the auto power-spectrum and cross-power spectrum of the  \ion{C}{ii} and \ion{H}{i} 21-cm signals from the EoR. Lets consider a given signal $S$ have spatial fluctuations as $\delta_S (\bm{x},z) = S(\bm{x},z)-\Bar{S}(z)$. If the fluctuations in the Fourier-domain is expressed as  $\Tilde{\delta}_S(\bm{k},z)$\footnote{This convention assumes a  definition of Fourier transform as\\ $\delta_S(\bm{x},z) = \int \frac{d^3k}{(2\pi)^3}\exp{(i\bm{k}\cdot\bm{x})} \Tilde{\delta}_S (\bm{k},z)$}, then the power spectrum of the signal $P_S(k)$ will be defined as,
\begin{equation}
    \langle \Tilde{\delta}_S^{*}(\bm{k}^{\prime}) \Tilde{\delta}_S(\bm{k}) \rangle = V\delta^K_{\bm{k}^{\prime}-\bm{k},0} P_S(k)\,,
    \label{eq:7}
\end{equation}
where $V$ denotes the volume under consideration, whereas $\langle...\rangle$ denotes the ensemble average. We have dropped the explicit dependence of $P(k)$ on $z$ in this notation, and it is assumed implicitly hereafter. We use the dimensionless power spectrum, $\Delta^2_S(k) = k^3P_S(k)/(2\pi^2)$ in our results to show the impact of light-cone effect.

The cross-power spectrum between two signals, $S1$ and $S2$, can be analogously defined as
\begin{equation}
    \langle \Tilde{\delta}^{*}_{S1}(\bm{k^{\prime}})\Tilde{\delta}_{S2}(\bm{k}) \rangle = V\delta^K_{\bm{k^{\prime}}-\bm{k},0}P_{S1\times S2}(k)
    \label{eq:8}
\end{equation}
The cross-correlation coefficient for these two signals is then expressed as
\begin{equation}
    r_{S1\times S2}(k)=\frac{P_{S1\times S2}(k)}{\sqrt{P_{S1}(k)P_{S2}(k)}}
\end{equation}
We discuss the impact of light-cone effect on $\Delta^2_{\text{\ion{C}{ii}}}(k)$, $\Delta^2_{\text{21cm}}(k)$ and $\Delta^2_{\text{\ion{C}{ii}$\times$21cm}}(k)$ for \ion{C}{ii} and \ion{H}{i} 21-cm signal from the EoR in the subsequent sections.

\subsection[CII power spectrum]{C\,{\sevensize II} power spectrum}
We compute the \ion{C}{ii} power spectrum for both coeval and light-cone cubes. The coeval power spectrum is first compared with \citet[]{Silva_2015} for consistency check. To have a quantitative estimate on how they compare, we have fitted a power law of the form $\Delta^2(k)=Ak^{n}$ to both \citet[]{Silva_2015} and our results. We find, for the \citet[]{Silva_2015}  $\textit{m1}$ model, that $A=3.86\times10^6$ and $n=1.67$, whereas our work results in a $A=1.41\times10^7$ and $n=1.74$. The relative difference in the spectral index in these two results is around $\sim 4\%$. This difference is most likely due to the differences in the population abundance  of \ion{C}{ii} emitters in these two works. Since the number of low-mass halos is high, a higher number of low-mass \ion{C}{ii} emitters would contribute more power to the larger $k$-modes than higher-mass halos to the lower $k$-modes. It means that we would expect a steeper slope (higher $n$) in our results. The higher population abundance in our case also contributes to a higher magnitude of the power spectrum.

\begin{figure}
    \centering
    \includegraphics[width=\columnwidth]{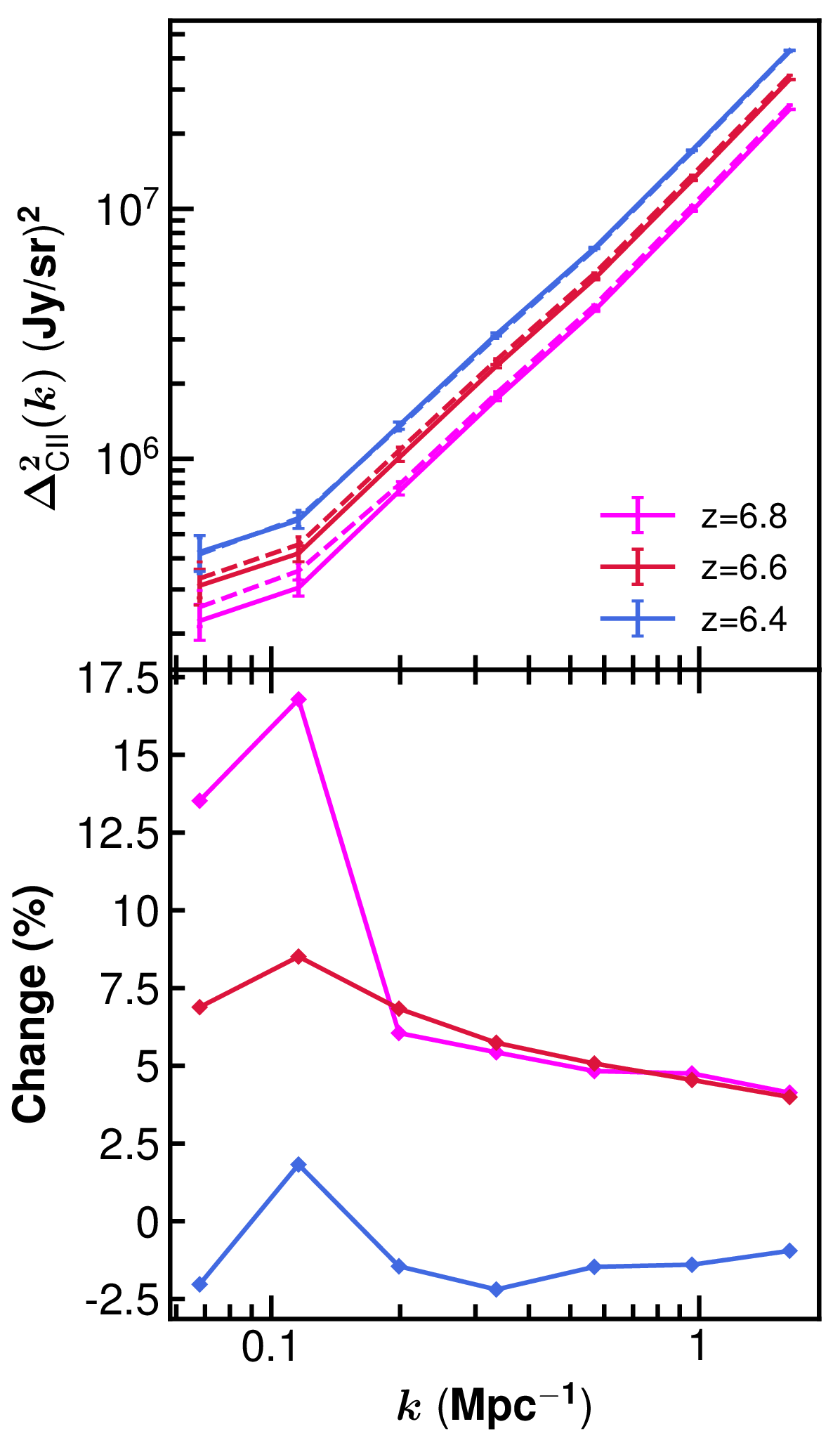}
    \caption{\ion{C}{ii} power-spectrum is shown in the top panel (solid lines represent coeval and dashed lines represent light-cone power-spectrum). Bottom panel shows the percentage change in power-spectrum due to light-cone effect.}
    \label{fig:CII_pwr}
\end{figure}

We next discuss the impact of the light-cone effect on the \ion{C}{ii} power spectrum. Fig.~\ref{fig:CII_pwr} shows the \ion{C}{ii} power spectrum (top panel) and the percentage change in the light-cone effect  with respect to the coeval power spectrum (bottom panel). The error bars in the power spectra are estimated from the sample variance in each $k$-mode bin. We compared the power spectra at three redshifts, $z=6.8, 6.6, \text{and}\, 6.4$. Our analysis is limited by the LoS extent of the light-cone volume, in this case it is such that we could not go beyond $z=6.4\, \text{and}\, 6.8$, although it is important to look at a few more redshifts beyond this limit. While comparing the light-cone power spectrum with the coeval power spectrum we choose the redshift of the centre of the light-cone chunk as our reference coeval redshift.

We observe that at $z=6.6\, \text{and}\, 6.8$, the change in the power spectrum is very much the same across all $k$-modes, except for $k \lesssim 0.1\, \text{Mpc}^{-1}$. At $k\sim 0.1\, \text{Mpc}^{-1}$, the light-cone effect is the most pronounced compared to the other $k$-modes. This is expected as the light-cone effect is a large scale effect (i.e. small $k$ effect in Fourier space). This can be understood in the following manner. To estimate the spherically averaged power spectrum of a signal if one increases the extent of the observed volume along the LoS (i.e. probing larger length scales along the LoS), one would be effectively combining signals from more distinctly different cosmic times, thus will see more pronounced impact of the light-cone effect in small $k$ power spectrum. 
 
Therefore, at larger scales, the behaviour of the light-cone effect becomes interesting and needs to be explored. Our coeval volume is limited to $(215\, \rm{Mpc})^3$, and to investigate the light-cone effect on the \ion{C}{ii} power spectrum for $k$-modes well below $\sim 0.1\, \text{Mpc}^{-1}$ with higher statistical significance, we will need to simulate the signal at even larger comoving volumes.

We observe that there is a systematic behaviour in the change of light-cone effect at $k\sim 0.1\, \text{Mpc}^{-1}$ (Fig.~\ref{fig:CII_pwr}, bottom panel). The impact of light-cone effect at this $k$-mode drops with decreasing redshifts. The maximum change in power due to the light-cone exceeds $15\%$ at $z=6.8$ at this $k$-mode; this shows that the impact of light-cone on the \ion{C}{ii} power spectrum is significant.

The light-cone cubes are labelled with $z_\text{c}$, which is the redshift at the centre of that volume. As discussed earlier, we compare the power spectrum of this cube with a coeval cube of redshift $z_\text{c}$. The properties of the fluctuations in the \ion{C}{ii} signal will be different in the central part of the light-cone cube compared to the front (lower $z$) and the far (higher $z$) end. The \ion{C}{ii} LIM signal essentially follow a biased  distribution of underlying dark matter density distribution, thus the signal coming from the far end of the light-cone volume will have less fluctuations than the front end (as structures are more clustered). Thus the relative difference in the power spectrum amplitude between $z_\text{c}$ (central portion) and these sections will be different. When $z_\text{c}$ is lower, these relative difference drop by unequal amounts. Compared to the far end of the light-cone volume, the power spectrum from the front part becomes more saturated (as formation and clustering of structures also saturates and overall evolution of structures slow down). Thus, the relative difference in power from the front portion and the central portion drops more, and we see that the impact of the light-cone effect drops with decreasing $z_\text{c}$.

\subsection[HI power spectrum]{H\,{\sevensize I} power spectrum}
Our main focus of this article is to quantify the light-cone effect on the \ion{C}{ii} power spectrum and the \ion{C}{ii}$\times$21-cm cross-power spectrum. However, we discuss the light-cone effect on the \ion{H}{i} 21-cm power spectrum, which has been studied extensively earlier, in this section for consistency checks and also as a prerequisite to the cross-power spectrum. Unlike the \ion{C}{ii} power spectrum, the 21-cm power spectrum is dependent on the state of the IGM or the reionization history, as the 21-cm signal fluctuations are mostly driven by the size and distribution of the ionized regions in the IGM. Thus one would expect that the impact of the light-cone effect will be different in case of this signal compared to the \ion{C}{ii} signal. The impact of the light-cone effect will vary depending on the reionization history. A faster reionization history is expected to exhibit a more drastic impact due to the light-cone effect.

\begin{figure*}
    \centering
    \includegraphics[width=\textwidth]{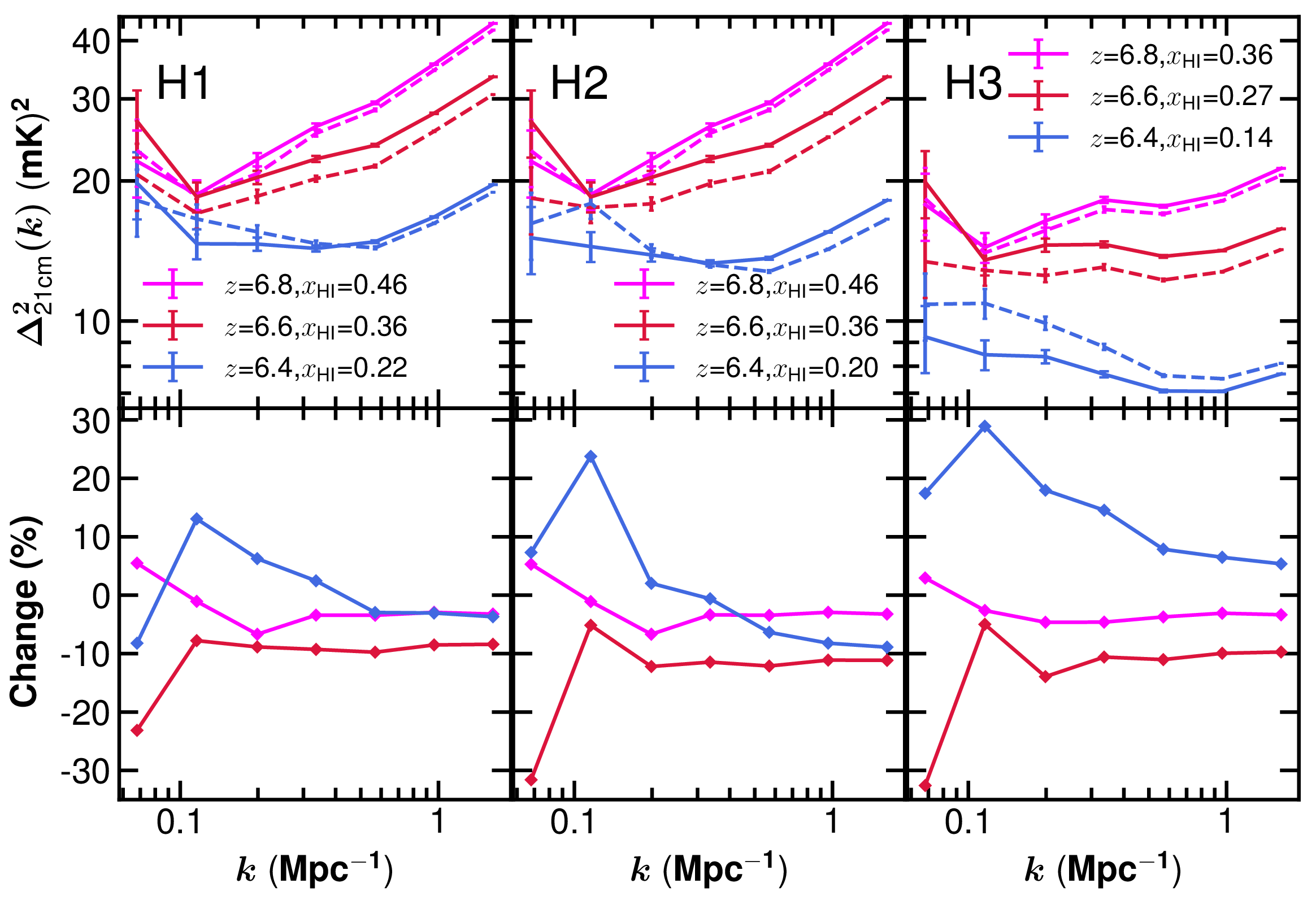}
    \caption{Top panel shows the coeval (solid) and light-cone (dashed) \ion{H}{i} 21-cm power spectrum for different neutral fractions ($\Bar{x}_{\text{\ion{H}{i}}}$). Bottom panel shows the impact of light-cone on the power spectrum in percentage. Panels from left to right (H1,H2 and H3) correspond to the different reionization histories.}
    \label{fig:HI_pwr}
\end{figure*}

The top panel of Fig.~\ref{fig:HI_pwr} shows the coeval and light-cone power spectrum of the \ion{H}{i} 21-cm signal. The bottom panel shows the impact of the light-cone effect in percentage. The \ion{H}{i} 21-cm power spectrum does not have a constant slope in the $\Delta^2(k)-k$ space, unlike the \ion{C}{ii} power spectrum, which is almost featureless. We find that the 21-cm power spectrum in our work is consistent with \citet[]{Datta_2012} but not exactly the same. The differences in the power spectra can be ascribed to the difference in the reionization history, source model and the method of simulating the EoR 21-cm signal in \citet[]{Datta_2012} and this work. In the simulations used by \citet[]{Datta_2012}, the reionization ends before $z=8$ and these simulations are based on full radiative-transfer technique. Even if we compare our 21-cm power spectra at the level of same mass-averaged IGM neutral fraction with that of the \citet[]{Datta_2012}, we still find a significant difference in their amplitude. This is due to the fact that as in \citet[]{Datta_2012} the same IGM neutral fraction is reached at a higher redshift thus the nature of the ionizing sources are different than that of our simulations. This is because the source model in both simulations are strongly dependent on the underlying halo mass function which will be significantly different at two far apart cosmological redshifts.  

The light-cone \ion{H}{i} 21-cm power spectrum varies across the three reionization histories that we have considered here. At $z=6.6$, which roughly represents the redshift mid-point for all three reionization histories, the light-cone power spectrum has different amplitudes for reionization histories H1 and H2. We observe that 21-cm power spectrum for H2 is more suppressed (Fig.~\ref{fig:HI_pwr}, top panels) compared to that of H1, though they have the same mass averaged neutral fraction at this redshift. The amplitude of the H3 light-cone power spectrum is suppressed at this redshift at a level similar to H2, for all $k$-modes except at $k \sim 0.2\, \text{Mpc}^{-1}$. If we compare the light-cone power spectra in these three reionization histories with their respective coeval power spectra, we see an overall decrease in amplitude due to the light-cone effect by approximately $\approx 10\%$ or more in all $k$-modes for $z =  6.6$. As one goes to the lower redshifts e.g. $z=6.4$, we notice a change in the nature of the light-cone effect. The light-cone power spectrum amplitude gets boosted compared to the coeval power spectrum specifically for large length scales i.e. small $k$-modes. This impact is maximum for reionization history H3. For the H3 history the power spectrum amplitude gets a boosting of around $\approx 20\%$ or more. This is consistent with \citet[]{Datta_2012}.    

As the light-cone effect in the EoR 21-cm signal is dependent on the reionization history, thus we observe a stronger impact when the neutral fraction evolves more rapidly with redshift e.g. the history H3. Since the \ion{C}{ii}$\times$21-cm cross-power spectrum has a contribution from the 21-cm signal fluctuations, thus we expect some of the behaviour  to reflect in the cross power spectrum as well, e.g. the variation of the impact of the light-cone with reionization history.

\subsection{CII\texorpdfstring{$\times$}{}21cm cross-power spectrum}
\begin{figure*}
    \centering
    \includegraphics[width=\textwidth]{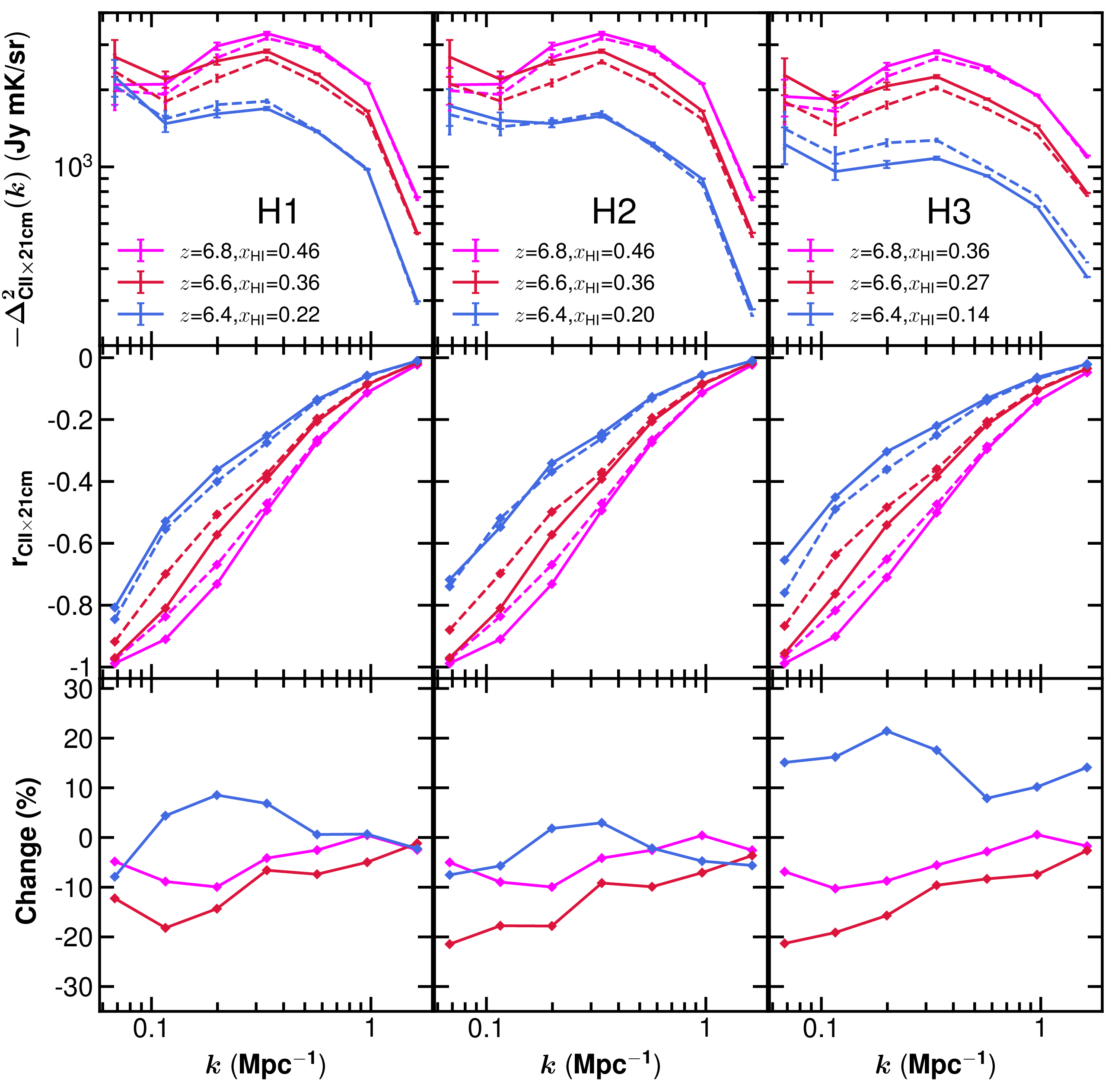}
    \caption{The \ion{C}{ii}$\times$21cm cross-power spectrum (top panel), along with cross-correlation coefficient (middle panel) and change in cross power due to light-cone effect (bottom panel) is shown here. Panels from left to right (H1, H2 and H3) correspond to different reionization histories. In the top and middle panel, solid and dashed lines represent coeval and light-cone cross-power spectrum respectively.}
    \label{fig:cross_power}
\end{figure*}

One of the main focuses of this work is to study the impact of light-cone effect on the \ion{C}{ii}$\times$21-cm cross-power spectrum. Both the \ion{C}{ii} and \ion{H}{i} 21-cm signal fluctuations contribute to the cross-power spectrum. It implies that the impact of the light-cone effect on the cross-power spectrum to vary with the reionization history under consideration, as it does in case of the \ion{H}{i} 21-cm power spectrum. Fig.~\ref{fig:cross_power} shows the \ion{C}{ii}$\times$21-cm cross-power spectrum, along with the cross-correlation coefficient and the percentage change in  the light-cone cross power spectrum with respect to the coeval cross-power spectrum.

During the EoR, most of the neutral hydrogen lives in the form of diffused gas within the IGM. The fluctuations in the \ion{H}{i} 21-cm signal will thus trace the fluctuations in the neutral gas density. In a typical inside-out reionization scenario (as in case of our simulations here), the ionizing photons released by the reionization sources first ionizes their local IGM and then spreads out to far away regions. As discussed earlier we would expect the \ion{C}{ii} LIM observations to probe the distribution and clustering pattern of these ionizing sources. During the early and middle stages of the EoR, we would expect the ionized regions to form around the sources of ionization and there will be no 21-cm signal. This implies that we should expect a strong anti-correlation between the \ion{C}{ii} intensity maps (tracing the sources) and \ion{H}{i} 21-cm maps (tracing the neutral hydrogen) at these stages. A visual inspection of the simulated light-cone maps of 21-cm and \ion{C}{ii} signal shown in Fig.~\ref{fig:lc} confirms this expectation.  It is quite apparent from this figure that most of the \ion{C}{ii} emission are coming from locations where the IGM is ionized or there is no 21-cm signal. We next inspect the cross-power spectrum and the cross-correlation coefficients shown in Fig.~\ref{fig:cross_power}. For all three reionization histories considered here there is a strong anti-correlation, marked by the negative value of the cross-power spectrum and the cross-correlation coefficient (see top and middle panels of Fig.~\ref{fig:cross_power}) between these two signals in the $k$-mode range $0.07 \leq k \leq 0.5 \, \text{Mpc}^{-1}$, when the reionization is half-way through i.e. $z = 6.8$. The degree of anti-correlation gradually decreases with the progress of reionization for all three reionization histories.

The impact of the light-cone effect on the cross-power spectrum at higher redshifts ($z = 6.8$) i.e. during the early and middle stages of the EoR is similar to that on the 21-cm power spectrum. There is an overall suppression in the cross-power spectrum amplitude during these stages of the EoR for all three reionization histories. The level of suppression in the cross-power amplitude is $\sim 10\%$ or less for smaller $k$-modes. The light-cone effect effectively reduces the amount of anti-correlation between these two signals at this stage. At the later stages of the EoR ($z = 6.6$) the light-cone effect continues to suppress the cross-power spectrum amplitude further. For histories H2 and H3, where reionization runs faster at later stages, the suppression in the cross-power due to the light-cone effect at large length scales can be as large as $\sim 20\%$. The impact on the H1 case for this statistic at this stage is comparatively less relative to H2 and H3. During the very late stages of the EoR ($z = 6.4$) the nature of impact due to the light-cone effect on the cross-power spectrum changes and instead of suppressing the power it enhances amplitude of the cross-power spectrum significantly. The impact is maximum in case of reionization history H3. We observe more than $20\%$ enhancement in cross-power for smaller $k$-modes in this scenario. In case of H1, the enhancement in power for small $k$-modes lies somewhere in between $5-10\%$, whereas for H2 it is less than or equal to $\sim 5\%$. These difference in the impact of light-cone effect can be ascribed to the nature of evolution of the reionization history in these three cases.

\section{Summary} \label{Summary}
In this work, we have investigated the impact of the light-cone effect on the auto and cross-power spectra of the EoR \ion{C}{ii} and 21-cm line intensity mapping signals. Using an N-body simulation, an FoF halo finding scheme and semi-numerical schemes to generate \ion{C}{ii} and \ion{H}{i} 21-cm intensity maps, we have built the light-cone cuboids for \ion{C}{ii} and 21-cm signals from the EoR. By performing a relative comparison between the auto and cross-power spectra estimated from the coeval and light-cone maps of these two signals, we have quantified the impact of the light-cone effect on these statistics. As per our knowledge this is the first effort to quantify the impact of the light-cone effect on the \ion{C}{ii} power spectrum and the \ion{C}{ii}$\times$21-cm cross-power spectrum from the EoR. We summarise the key findings of our work below:
\begin{itemize}
    \item The light-cone effect has a significant impact on the \ion{C}{ii} power spectrum, and it is most pronounced at large length scales i.e. $k\sim 0.1\, \text{Mpc}^{-1}$. This effect can introduce a change in power by $\sim 15\%$ (at $z=6.8$).
    \item We find that impact of light-cone effect on the \ion{C}{ii} power spectrum decreases with decreasing redshift and by $z=6.4$ it  leads to a suppression of the amplitude of the \ion{C}{ii} power spectrum in most of the $k$-modes by a small amount ($\sim 2\%$).
    \item The reduction in the impact of the light-cone effect on the \ion{C}{ii} power spectrum with decreasing redshift can be understood in the following manner: The \ion{C}{ii} power spectrum follows the underlying halo distribution which is a manifestation of the underlying structures. The rate of growth of structures is relatively rapid with redshift at earlier cosmic times. This growth rate slows down significantly as we approach the lower redshifts. Thus the difference in structures between a coeval map and a light-cone map at lower redshifts is significantly small, thus the impact of the light-cone effect on the \ion{C}{ii} power spectrum is also relatively small when compared with the same at higher redshifts.
    \item The impact of the light-cone effect on the \ion{C}{ii}$\times$21-cm cross-power spectrum is significant, with the maximum change in cross-power amplitude reaching up to 20\%.
    \item For the \ion{C}{ii}$\times$21-cm cross-power spectrum, the impact of light-cone effect strongly depends on the reionization history. 
    A rapid evolution in the mass-averaged neutral fraction with redshift results in a stronger light-cone effect on the cross-power spectrum. 
    \item The nature of the impact of the light-cone effect, translated in terms of either suppression or enhancement of the cross-power spectrum amplitude depends on the stage of reionization that the light-cone volume under consideration is capturing. If the light-cone volume is sampled from the early or middle stages of the EoR, it results in a suppression of the cross-power amplitude. However, when the light-cone volume represents very late stage of the EoR, it results in an enhancement of the cross-power amplitude.
\end{itemize}

Our analysis in this article suggest that the light-cone effect has a significant impact on the amplitude of the \ion{C}{ii} intensity map power spectra as well the \ion{C}{ii}$\times$21-cm cross-power spectrum from the EoR. It is therefore essential to take into account this effect while modeling the signal and estimating parameters of reionization from observations.

In our model of \ion{C}{ii} emission used here, we have not taken into account the population abundance of \ion{C}{ii} emitters. We have assumed that every halo is an efficient emitter of the \ion{C}{ii} line. When compared with \citet[]{Silva_2015}, this leads to a different estimation of the \ion{C}{ii} power spectrum. However, we expect this to have a very little effect on the results described above.

The CII luminosity-SFR mapping prescription (equation \eqref{eq:6}) adopted from \citet{Silva_2015} does not have any redshift dependence. If one chooses a different model where this mapping does have a redshift dependence, then the impact of light-cone on the observed \ion{C}{ii} statistics will also change.

The reionization models that we have considered here do not take into account the effects of density dependent hydrogen recombination. If properly taken into account, it is expected to have an impact on both the reionization history and the 21-cm topology, which in turn will affect the light-cone effect on the \ion{C}{ii}$\times$21-cm cross-power spectrum.

In this work, we have only focused on the impact of light-cone on the observable statistics of the signal. The systematics (instrumental effects, modelling the foregrounds and impact of the interlopers) for the \ion{C}{ii} signal and the light-cone effect are two important and independent issues that we need to model separately. We need to address both issues for accurate prediction of the signal. In particular, the \ion{C}{ii} line emission will be contaminated from interlopers such as CO line emissions from comparatively low-redshift galaxies \citep[][]{Gong_2011,Silva_2015,Sun_2018}. In this study, we have assumed that the foreground emissions are perfectly removed from both the \ion{C}{ii} as well as the 21-cm data. However, in reality, this might be a difficult goal to achieve. This in principle will increase the errors in the \ion{C}{ii} power spectrum and the \ion{C}{ii}$\times$21-cm cross-power spectrum.

Finally, our present simulation box size is limited to $215$ Mpc. A study of the light-cone effect at $k \leq 0.1\, \text{Mpc}^{-1}$ with improved statistical significance will require a larger box size. As we have demonstrated here that the light-cone effect is mostly a large scale effect. Thus studying the power spectrum and cross-power spectrum for those $k$-modes will be worthwhile. Analytical modelling of the light-cone effect is important, which might allow us to assess this impact quickly. However, so far, no attempts have been made in this direction. We would like to address many of the issues mentioned above in our future follow-up works.

\section*{Acknowledgements}

The authors would like to thank the anonymous referee for their encouraging comments and useful feedback. CSM would like to acknowledge funding from the Council of Scientific and Industrial Research (CSIR) via a CSIR-JRF fellowship for this project, under the grant 09/1022(0080)/2019-EMR-I. KKD acknowledges financial support from BRNS through a project grant (sanction no: 57/14/10/2019-BRNS). All of the simulations and the entire numerical analysis of the simulated data were done using the computing resources available to the Cosmology with Statistical Inference (CSI) research group at IIT Indore.

This research made use of: \texttt{matplotlib} \citep[]{Hunter:2007}, \texttt{numpy} \citep[]{harris2020array}, arXiv \footnote{\url{https://arxiv.org}} research-sharing platform , NASA Astrophysics Data System Bibliographic Services and Crossref \footnote{\url{https://www.crossref.org}} citation services.

%%%%%%%%%%%%%%%%%%%%%%%%%%%%%%%%%%%%%%%%%%%%%%%%%%
\section*{Data Availability}

The simulated data underlying this work will be shared upon reasonable request to the corresponding author.

%%%%%%%%%%%%%%%%%%%% REFERENCES %%%%%%%%%%%%%%%%%%

% The best way to enter references is to use BibTeX:

%\nocite{*}
\bibliographystyle{mnras}
\bibliography{references}

%%%%%%%%%%%%%%%%% APPENDICES %%%%%%%%%%%%%%%%%%%%%

%\appendix

%%%%%%%%%%%%%%%%%%%%%%%%%%%%%%%%%%%%%%%%%%%%%%%%%%

% Don't change these lines
\bsp	% typesetting comment
\label{lastpage}
\end{document}